\documentclass[
    ,final            
  ]
  {aipproc}

\usepackage{multirow}					

\layoutstyle{6x9}
\usepackage{amssymb,amsmath}

\begin{document}

\title{First Measurement of the Beam Normal Single Spin Asymmetry in $\Delta$ Resonance Production by $Q_{\rm weak}$}


\author{Nuruzzaman for the $Q_{\rm weak}$ Collaboration\index{Nuruzzaman}}{
  address={Hampton University, Hampton, VA 23668, USA.}
}
\begin{abstract}
The beam normal single spin asymmetry ($B_{\rm n}$) is generated in the scattering of transversely polarized electrons from unpolarized nuclei. The asymmetry arises from the interference of the imaginary part of the two-photon exchange with the one-photon exchange amplitude. The $Q_{\rm weak}$ experiment has made the first measurement of $B_{\rm n}$ in the production of the $\Delta$(1232) resonance, using the $Q_{\rm weak}$ apparatus in Hall-C at the Thomas Jefferson National Accelerator Facility. 
The final transverse asymmetry, corrected for backgrounds and beam polarization, is $B_{\rm n}$ = 43~$\pm$~16~ppm at beam energy 1.16~GeV at an average scattering angle of about 8.3~degrees, and invariant mass of 1.2~GeV.
The measured preliminary $B_{\rm n}$ agrees with a preliminary theoretical calculation. 
$B_{\rm n}$ for the $\Delta$ is the only known observable that is sensitive to the $\Delta$ elastic form-factors ($\gamma$*$\Delta\Delta$) in addition to the generally studied transition form-factors ($\gamma$*N$\Delta$), but extracting this information will require significant theoretical input. 

\end{abstract}

\maketitle


\section{Beam Normal Single Spin Asymmetry}
\label{Beam Normal Single Spin Asymmetry}


The beam normal single spin asymmetry ($B_{\rm n}$) is generated in the scattering of transversely polarized electrons from unpolarized nuclei. The asymmetry arises from the interference of the imaginary part of the two-photon exchange with the one-photon exchange amplitude in electron-nucleon scattering. $B_{\rm n}$ is a parity-conserving asymmetry and is time-reversal invariant and can be expressed as~\cite{DeRújula1971365, Gorchtein2005273} 

\begin{equation}
	\label{eq:B_n_asym}
	B_{\rm n}(\phi) = \frac{\sigma\uparrow - \sigma\downarrow}{\sigma\uparrow + \sigma\downarrow} = \frac{2~Im T_{2\gamma}\bullet T_{1\gamma}^{*} }{|T_{1\gamma}|^2}.
\end{equation}

\begin{figure}[!h]
  \includegraphics[width=3.8in]{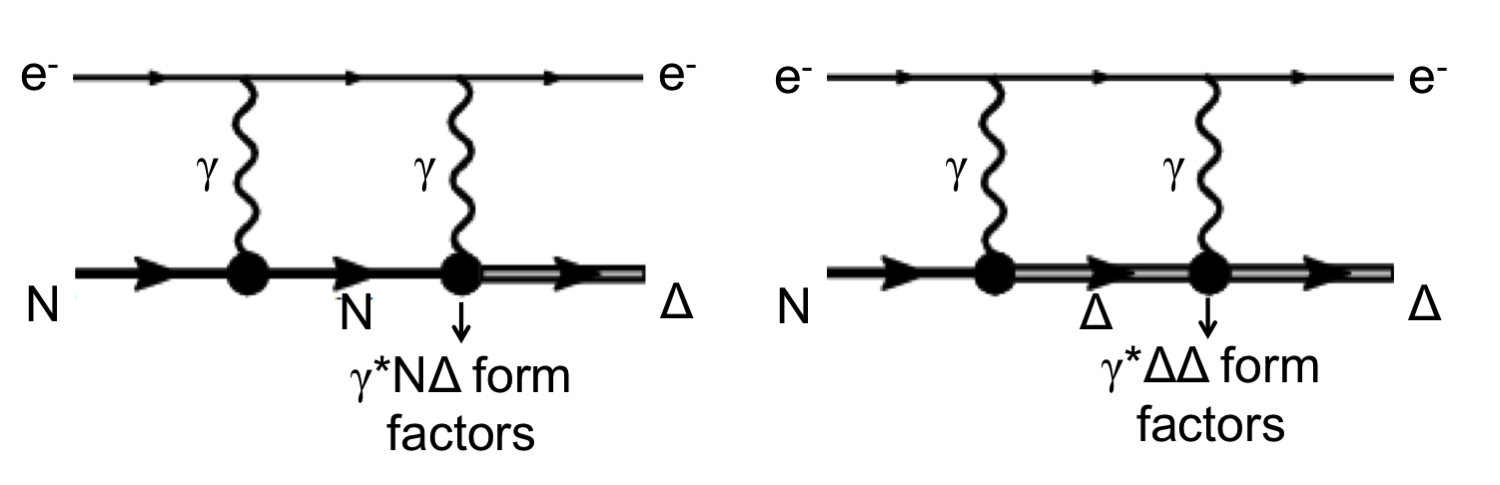}
  \caption{The beam normal single spin asymmetry in inelastic electron-nucleon scattering with the $\Delta$ in the final state of the two-photon exchange process. The nucleon as the intermediate state is shown in the left, whereas $\Delta$ as intermediate state is shown in the right. The knowledge on the $\gamma\Delta\Delta$ vertex is scarce.}
  \label{fig:two-photon_exchange}
\end{figure}

Here $\sigma\uparrow(\downarrow)$ is the cross section for spin up (down) electrons and $T_{1\gamma}$($T_{2\gamma}$) is the amplitude for the 1-photon (2-photon) exchange. $B_{\rm n}$ vanishes in the Born approximation, and therefore provides a direct access to the imaginary part of the two-photon exchange amplitude. 
The imaginary part of the two photon exchange contains information about the nucleon intermediate states (see FIGURE~\ref{fig:two-photon_exchange}). 
The proton and $\Delta$ intermediate hadron vertices are known except for the $\gamma$*$\Delta\Delta$ electromagnetic (EM) vertex. The proton EM form factors~\cite{Perdrisat2007694} and $N\rightarrow\Delta$ EM transition form factors~\cite{PhysRevC.88.032201, Segovia} have been  investigated in the last decade. 
The electron-proton scattering with $\Delta$ in the final state gives an unique opportunity to study the $\gamma$*$\Delta\Delta$ form factor~\cite{Segovia}. 
The exploration of the $\gamma$*$\Delta\Delta$ form factor has potential to constrain the nuclear properties of the $\Delta$, such as charge radius and magnetic moment~\cite{presentation:mark}. 
The $Q_{\rm weak}$ experiment~\cite{qweak_proposal_2007} has made the first measurement of $B_{\rm n}$ in the production of the $\Delta$(1232) resonance in electron-proton scattering, using the $Q_{\rm weak}$ apparatus in Hall-C at the Thomas Jefferson National Accelerator Facility. 

\begin{figure}[!h]
  \includegraphics[width=5.8in]{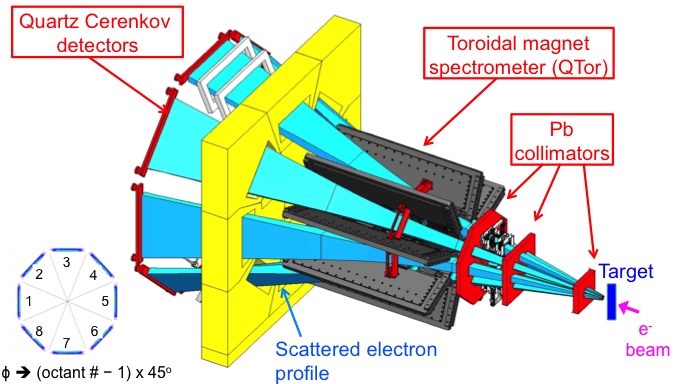}
  \caption{(Color online) The experimental design showing the target, collimation, toroidal magnetic spectrometer (QTor), electron trajectories, and the \v{C}erenkov detectors. The QTor magnetic spectrometer focuses inelastically scattered electrons (blue profile) onto the eight main detector bars azimuthally oriented around the beamline (detector coordinate system is shown in the left sub figure). The distance along the beam line from the target center to the center of the quartz \v{C}erenkov detector array is $\sim$12.2~m.}
  \label{fig:experimentalApparatus}
\end{figure}


\section{Experimental Measurement of the Asymmetry}
\label{Experimental Measurement of the Asymmetry}

The $Q_{\rm weak}$ longitudinal measurement setup~\cite{qweak_proposal_2007} was used for this measurement with minor modifications. The electron beam polarization was changed from the nominal longitudinal setup to produce fully transverse polarization, either horizontal or vertical relative to the accelerator bend plane, using the double Wien filter and solenoid magnets at the injector~\cite{Allison2015105}. The magnetic field of the torodial magnetic spectrometer (QTor) was lowered to 6700~A to focus the inelastically scattered electrons onto the main \v{C}erenkov detectors (see FIGURE~\ref{fig:experimentalApparatus}).

\begin{table}[!h]
\centering
   \caption
	{The data set for the $B_{\rm n}$ measurement in the production of the $\Delta$(1232) resonance. The data with vertical transverse polarization are in parentheses, the rest are for horizontal transverse polarization. The beam currents and total charge on target in Coulombs are shown in the table. More information about the conditions of data taking is given in~\cite{nur_qweak_thesis}. A full analysis of the data on H$_{2}$ target at the $\Delta$ peak, indicated by $\dagger$, is discussed in this article. The transverse asymmetry on an Al target, indicated by $\dagger\dagger$, was also analyzed as a background correction for the H$_{2}$ data set. The analysis of the remaining data are ongoing and will not be covered in this article.
	}
  \begin{tabular}{ c | c | c  c  c | c  c }
    \noalign{\hrule height 1pt}
    \multirow{3}{*}{} & \multicolumn{6}{c}{QTor current} \\ \cline{2-7}
		 &  6000 A & & 6700 A ($\Delta$ peak) & &  \multicolumn{2}{c}{7300 A}\\
	\cline{2-7}
	     & H$_{2}$ & H$^{\dagger}_{2}$ & Al$^{\dagger\dagger}$ & $^{12}$C &  H$_{2}$ & Al \\
    \noalign{\hrule height 1pt}
	   Collected Data [C] & 1.5 & 1.8 (1.9) & 0.8 (0.4) & 0.6 & 2.0 & 0.9 \\
    \noalign{\hrule height 1pt}
	   Beam current $I$ [$\mu$A] & 180 & 180 & 60 & 75 & 180 & 60 \\
    \noalign{\hrule height 1pt}
   \end{tabular}
 \label{tab:transverse_inelastic_data_set}
\end{table}


The total collected data after the hardware and software quality cuts is shown in TABLE~\ref{tab:transverse_inelastic_data_set}. The QTor current of 6700~A selects the inelastic events near the invariant mass,$W$, of 1.2~GeV. Data on both sides of the $\Delta$ peak (6000~A and 7300~A) were taken to better constrain the elastic dilution. 
Data were collected on a primary liquid hydrogen (H$_{2}$) cell, a 4\% thick downstream aluminum alloy (Al) to constrain the target window backgrounds, and a 1.6\% thick downstream carbon foil ($^{12}$C) to help understand the atomic number dependence of $B_{\rm n}$ at beam energy of 1.16~GeV for both spin orientations. 
The acceptable beam currents for the different targets were different (see TABLE~\ref{tab:transverse_inelastic_data_set}) as it passes through the different targets of different effective thicknesses. 
The beam was rastered on the target over an area of 4~mm$\times$4~mm by the fast raster system to minimize the target boiling and to prevent the beam from burning through the target windows.
The insertable half wave plate was used to help suppress the helicity correlated beam asymmetries and was reversed at intervals of about 2~hours. 


A single detector asymmetry was obtained by averaging the two PMT asymmetries from each \v{C}erenkov detector. The error weighted average of the asymmetries from runlets, about 5 minute long data samples, was extracted for a given data set. 
The raw asymmetries were corrected for the false asymmetries using linear regression as 
$\epsilon_{\rm reg} = \epsilon_{\rm raw} - \displaystyle\sum_{i=1}^{5}\frac{\partial \epsilon_{\rm raw}}{\partial T_{i}}\Delta T_{i}$, where $T_{i}$ represents different beam parameters.
To extract the regressed asymmetry $\epsilon_{\rm reg}$ from the detectors, the average asymmetry for the two different insertable half wave plate (IHWP) settings, IN and OUT, were determined separately for each main detector bar. The asymmetries measured in the IHWP configurations were sign corrected for the extra spin flip and averaged together after checking for the IHWP cancellation of the false asymmetries. The error weighted value of $<$IN,-OUT$>$ determined the measured regressed asymmetry for each bar. These regressed asymmetries were then plotted against the detector octant number, which represents the location of the detector in the azimuthal plane ($\phi$ = (octant - 1)$\times$45$^{\circ}$), and they were fitted using a function of the form $\epsilon_{\rm reg}^{\rm H}\sin(\phi)/\epsilon_{\rm reg}^{\rm V}\cos(\phi)$.
This analysis was focused on the azimuthal dependence of the detector asymmetries representing the transverse asymmetries.


\begin{figure}[!h]
  \includegraphics[width=5.8in]{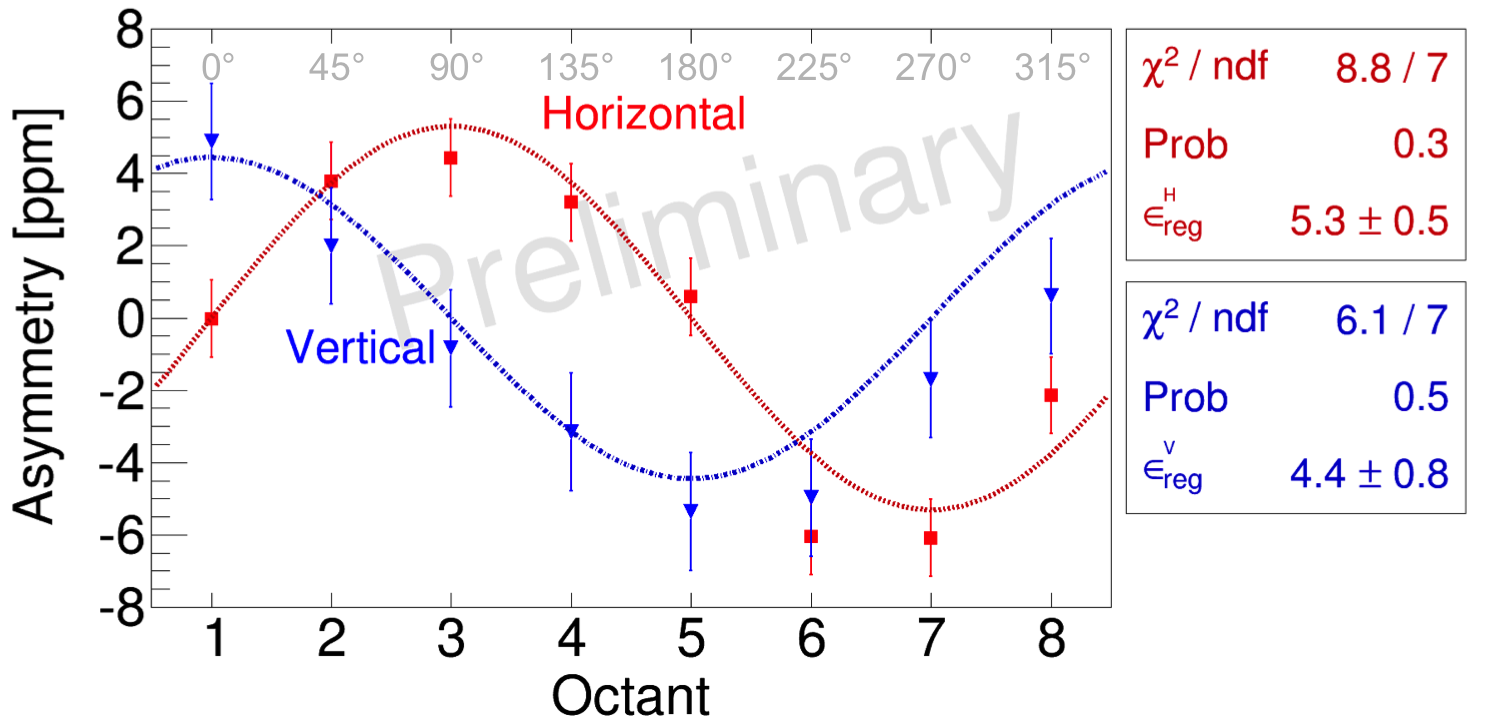}
  \caption{(Color online) The measured regressed main detector asymmetries for horizontal and vertical transverse polarization are shown with red squares and blue triangles, respectively. Data points for horizontal transverse are $\sim$4~hour long measurement, whereas vertical transverse data points are $\sim$2~hour long. The uncertainties are statistical only. The fit functions used are $\epsilon_{\rm reg}^{\rm H} \sin(\phi)$ for horizontal transverse and $\epsilon_{\rm reg}^{\rm V} \cos(\phi)$ for vertical transverse, respectively. Asymmetries in each case shows $\sim$90$^{\rm o}$ phase offset, as expected between horizontal and vertical configurations.}
  \label{fig:measuredAsymmetry}
\end{figure}

\begin{equation}
	\label{eq:B_n_measured_asym}
	\epsilon_{\rm reg}(\phi) = \frac{N\uparrow - N\downarrow}{N\uparrow + N\downarrow} = -B_{\rm n}\vec{S}\cdot\hat{n}=B_{\rm n}|S|\sin(\phi-\phi_{0}) = B_{\rm n}\left[P^{\rm V}\cos(\phi) - P^{\rm H}\sin(\phi)\right]
\end{equation}

Here, $\phi$ is the azimuthal angle in the transverse plane to the beam direction. $\phi$ = 0 indicates beam left, $\phi_{0}$ is a possible phase offset expected to be consistent with zero. 
$P^{H}$ and $P^{V}$ are horizontal and vertical components of the transverse polarization, respectively. $\epsilon_{\rm reg}$ is the measured regressed asymmetry (amplitude) of the azimuthal modulation generated by $B_{\rm n}$. 
%
The measured regressed asymmetries for the horizontal and vertical transverse polarization on H$_{2}$ target are $\epsilon_{\rm reg}^{H}$ = 5.3~$\pm$~0.5~ppm and $\epsilon_{\rm reg}^{V}$ = 4.4~$\pm$~0.8~ppm, respectively (see FIGURE~\ref{fig:measuredAsymmetry}). The error weighted asymmetry of the horizontal and vertical transverse data set is 5.1~$\pm$~0.4~ppm. The systematic uncertainty on the measured asymmetry has contributions from non-linearity, fit function, detector acceptance and other regression related uncertainties. The combined systematic uncertainty for this data set is about 0.1~ppm. The total uncertainty on the measured asymmetry is quadrature sum of the statistical and systematic uncertainties and is dominated by the statistical uncertainty.

\section{Extraction of Physics Asymmetry}
\label{Extraction of Physics Asymmetry}

The beam normal single spin asymmetry in inelastic electron-proton scattering is obtained from the measured asymmetry by accounting for the polarization, backgrounds, EM radiative corrections, and kinematics normalization using 

\begin{equation} \label{equ:PhysicsAsymmetry}
B_{\rm n} = M_{\rm kin} \left[ \frac{\left(\frac{\epsilon_{\rm reg}}{P}\right) - B_{\rm Al}f_{\rm Al} - B_{\rm BB}f_{\rm BB} - B_{\rm QTor}f_{\rm QTor} - B_{\rm el}f_{\rm el} }{1 - f_{\rm Al} - f_{\rm BB} - f_{\rm QTor} - f_{\rm el}} \right].
\end{equation}

Here $M_{\rm kin}$ is a correction factor for the experimental bias and radiative effects, $P$ is the beam polarization, and $B_{\rm bi}$ is $i^{th}$ background asymmetry with fraction of backgrounds in the total detector acceptance (dilution) $f_{\rm bi}$. 
$\epsilon_{\rm reg}$ is the regressed asymmetry after corrected for the false asymmetries using linear regression. 
The asymmetry was normalized with respect to beam polarization ($P$) using precision measurement of the polarization using the M{\o}ller polarimeter and Compton polarimeter. The measured polarization was $\sim$88\% with $<$1\% uncertainty.
The backgrounds were corrected for the aluminum window (Al), scattering from the beamline (BB), neutral particles in the magnet acceptance (QTor), and elastic radiative tail (el). Final correction was applied for the radiative tails and other kinematic correction ($M_{\rm kin}$). The background asymmetries and dilutions comes from the dedicated measurements and simulations. 

\begin{figure}[!h]
  \includegraphics[width=5.8in]{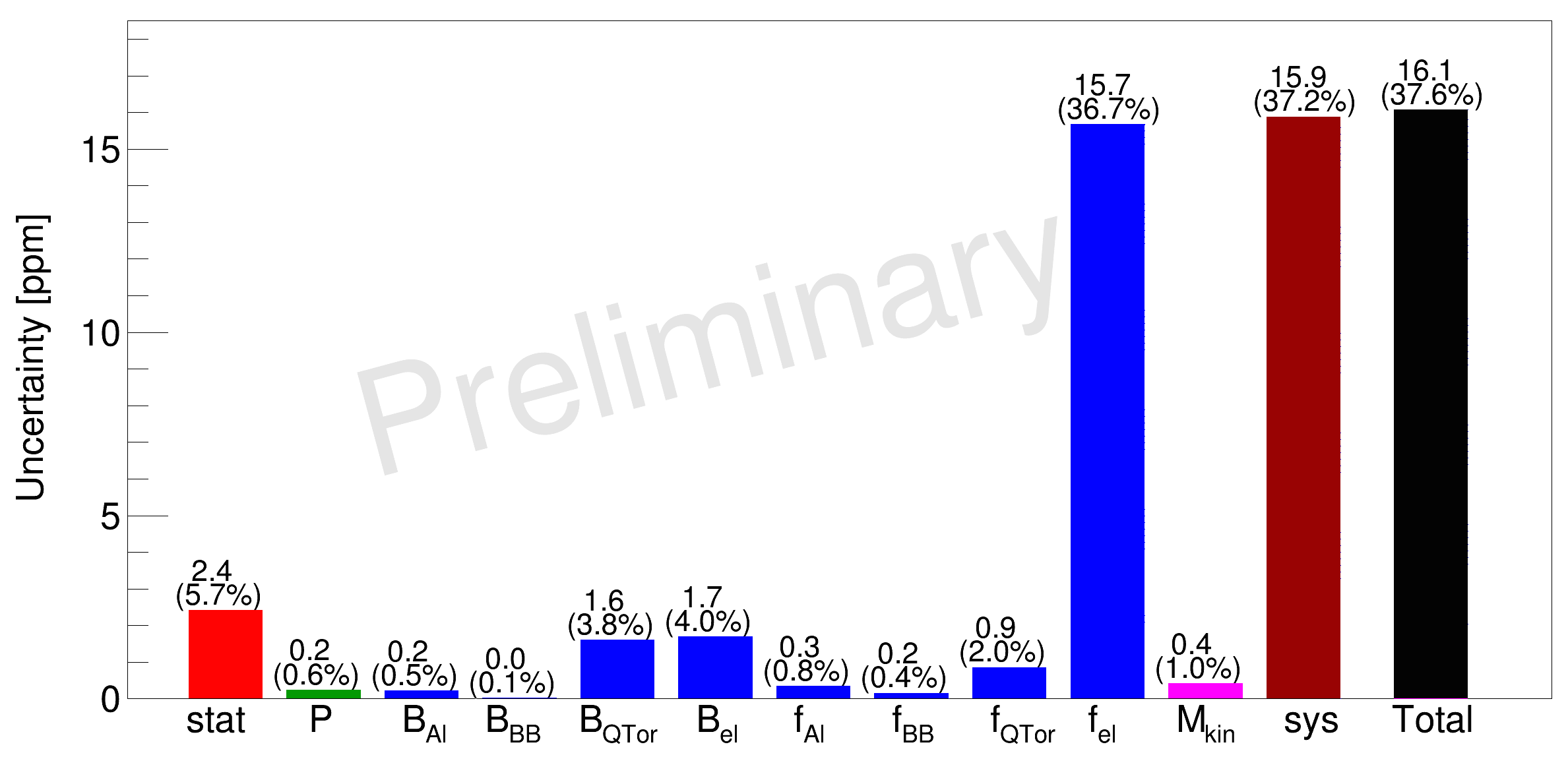}
  \caption{(Color online) Summary of the uncertainties on $B_{\rm n}$. The measured systematic contains uncertainties related to the extraction of the physics asymmetry such as regression, polarization, background, and kinematics. The total uncertainty is the quadrature sum of the statistical and systematic uncertainties. The total uncertainty is dominated by the systematic uncertainty from the elastic radiative tail. The uncertainties are in ppm and the corresponding relative uncertainties are shown in parentheses.}
  \label{fig:PhysicsUncertainty}
\end{figure}

\section{Preliminary Result}
\label{Preliminary Result}

Using all the input values in Equation~\ref{equ:PhysicsAsymmetry} gives the beam normal single spin asymmetry in inelastic electron-proton scattering

\begin{equation} \label{equ:FinalResult}
B_{\rm n} = 43 \pm 16~\text{ppm}
\end{equation}
for the effective kinematics of acceptance averaged electron energy $\langle E_{\rm beam}\rangle$ = 1.16~GeV, $\langle Q^{2}\rangle$ = 0.021~(GeV/c)$^{2}$, an average scattering angle $\langle\theta\rangle$ = 8.3$^{\rm o}$, and invariant mass $W$~$\sim$1.2~GeV. The contributions from the different uncertainty sources into the final measurement are summarized in 
FIGURE~\ref{fig:PhysicsUncertainty}. 
The dominant correction to the asymmetry comes from the elastic radiative tail whereas the dominant uncertainty on the measured asymmetry comes from statistics. The extraction of $B_{\rm n}$ depends strongly on the elastic dilution ($f_{\rm el}$). Careful study is ongoing to reduce the uncertainty in $f_{\rm el}$.

\begin{figure}[!h]
  \includegraphics[width=5.8in]{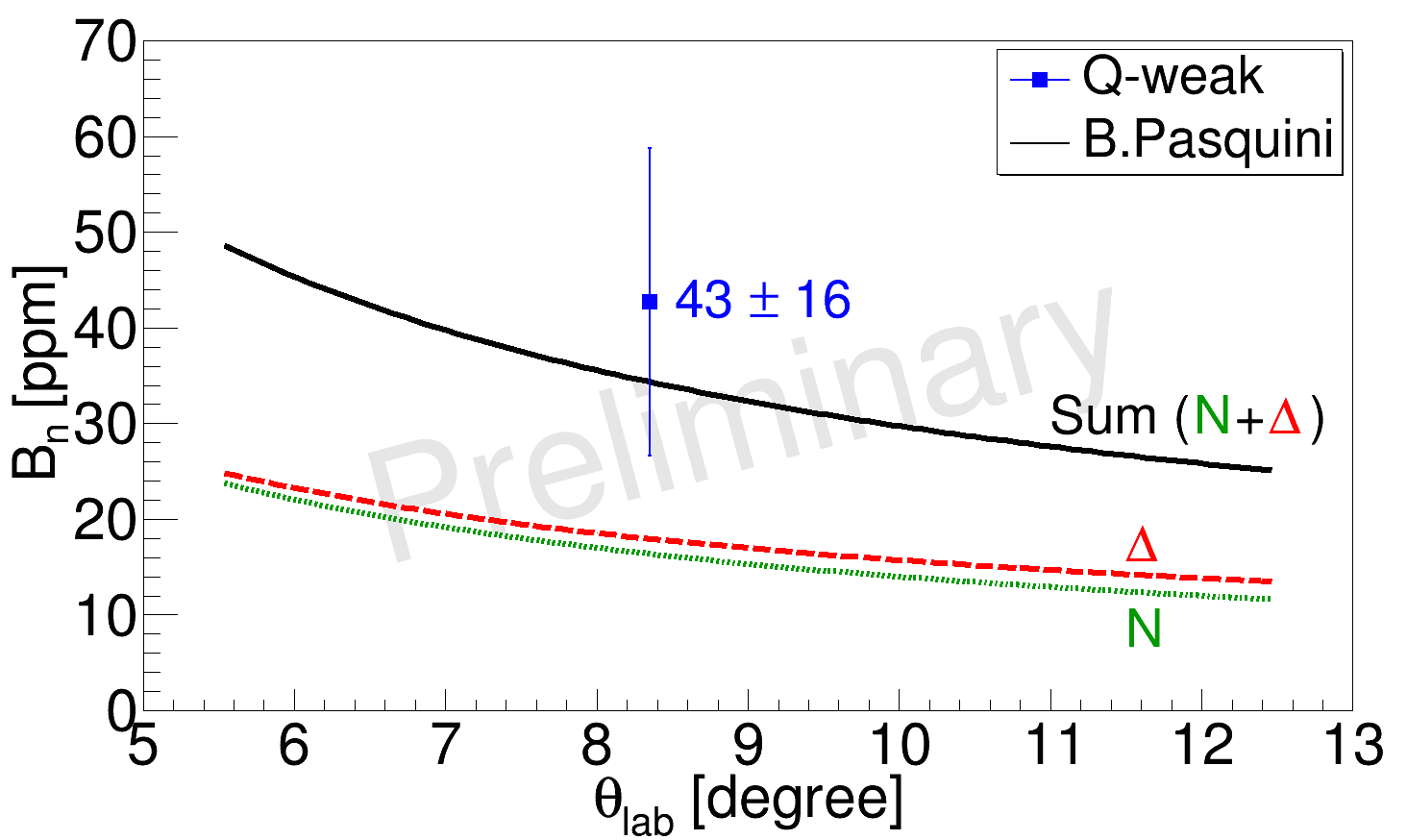}
  \caption{(Color online) Comparison of the $Q_{\rm weak}$ preliminary $B_{\rm n}$ measurement in inelastic electron-proton scattering at $\langle\theta_{\rm lab}\rangle$ = 8.3$^{\rm o}$ to the model calculations by B.Pasquini. The nucleon and $\Delta$ contributions are shown by dotted (green) and dashed (red) lines, respectively whereas, sum is shown by solid (black) line.}
  \label{fig:thetaDependenceOfAsymmetry}
\end{figure}

The measured $B_{\rm n}$ from this analysis is compared with a preliminary model calculation from B. Pasquini~\cite{barbara_communication}. The preliminary measured $B_{\rm n}$ agrees with the model calculation (see FIGURE~\ref{fig:thetaDependenceOfAsymmetry}). The theory asymmetry is large in the forward scattering angle region. The nucleon and $\Delta$ contribution in the asymmetry have similar magnitude and similar functional form. 
There is a lattice Quantum Chromo Dynamics parameterizations for the $\gamma$*$\Delta\Delta$ form factors in the model calculation which are not well understood. 
The physics implications of the model will be investigated with more detail in the near future. 
The $Q_{\rm weak}$ transverse data set has potential to constrain models and along with possible future world data has potential to study the charge radius and magnetic moment of $\Delta$.

In addition to this measurement, $Q_{\rm weak}$ performed beam normal single spin asymmetry measurements for various physics processes, which are summarized in TABLE~\ref{tab:transverse_data_set}. While these data sets are still being analyzed, their quality and relative precision makes them good candidates to test model predictions for the beam normal single spin asymmetries at forward angle kinematics from these different processes.

\begin{table}[!h]
\centering
   \caption
	{The full transverse data set by $Q_{\rm weak}$. A full analysis of the inelastic electron-proton scattering with a $\Delta$ in the final state at beam energy ($E_{beam}$) 1.16~GeV is presented in this article, indicated by $\ddagger$ (more details about this data set is discussed in TABLE~\ref{tab:transverse_inelastic_data_set}). The data were taken at forward scattering angle $\sim$8$^{\rm o}$ and at low $Q^{2}$ $\sim$0.02 - 0.03~(GeV/c)$^{2}$.
	}
  \begin{tabular}{ l | c }
    \noalign{\hrule height 1pt}
    Interaction & Target \\
    \noalign{\hrule height 1pt}
		Elastic e+p at $E_{beam}$ = 1.16~GeV~\cite{buddhini_cipanp2012} & H$_{2}$, Al, C \\
		Inelastic e+p with a $\Delta$ in the final state at $E_{beam}$ = 1.16~GeV$^{\ddagger}$ & H$_{2}$, Al, C \\
		Inelastic e+p with a $\Delta$ in the final state at $E_{beam}$ = 0.88~GeV & H$_{2}$ \\
		Elastic e+e at $E_{beam}$ = 0.88~GeV & H$_{2}$\\
		Inelastic e+p at W = 2.5~GeV & H$_{2}$\\
		Pion electro-production at $E_{beam}$ = 3.3~GeV & H$_{2}$\\
    \noalign{\hrule height 1pt}
   \end{tabular}
 \label{tab:transverse_data_set}
\end{table}

\section{Summary and Outlook}
\label{Summary and Outlook}

The $Q_{\rm weak}$ collaboration has made a $\sim$38\% relative measurement of the beam normal single spin asymmetry $B_{\rm n} = 43\pm~16~\text{ppm}$ using transversely polarized 1.16~GeV electrons scattering inelastically from the protons with an average $Q^{2}$ of 0.021~(GeV/c)$^{2}$, an average scattering angle of 8.3$^{\rm o}$, and invariant mass of 1.2~GeV. This is the first measurement of the beam normal single spin asymmetry in inelastic electron-proton scattering. The preliminary asymmetry agrees with the preliminary model calculation. The physics implications of the model calculation needs more investigation. The $Q_{\rm weak}$ transverse data set along with possible future world data has potential to constrain theoretical models and study the charge radius and magnetic moment of $\Delta$.

\begin{theacknowledgments}
This work was supported by DOE Contract No. DEAC05-06OR23177, under which Jefferson Science Associates, LLC operates Thomas Jefferson National Accelerator Facility. Special thanks to B. Pasquini for providing theory predictions at $Q_{\rm weak}$ kinematics.
\end{theacknowledgments}

\bibliographystyle{aipproc}   


\bibliography{nurTransverseN2Delta_CIPANP2015}

\IfFileExists{\jobname.bbl}{}
 {\typeout{}
  \typeout{******************************************}
  \typeout{** Please run "bibtex \jobname" to optain}
  \typeout{** the bibliography and then re-run LaTeX}
  \typeout{** twice to fix the references!}
  \typeout{******************************************}
  \typeout{}
 }

\end{document}